\documentclass[referee,sn-basic]{sn-jnl}
\jyear{2023}%

\theoremstyle{thmstyleone}%
\theoremstyle{thmstyletwo}%
\theoremstyle{thmstylethree}%

\begin{document}

\title[Platinum-based Catalysts for ORR
 simulated with a
Quantum Computer]{Platinum-based Catalysts for Oxygen
Reduction Reaction simulated with a
Quantum Computer}


\author*[1]{\fnm{Cono} \sur{Di Paola}}\email{cono.dipaola@quantinuum.com}
\author*[1] {\fnm{Evgeny} \sur{Plekhanov}}\email{evgeny.plekhanov@quantinuum.com}
\author[1]{\fnm{Michal} \sur{Krompiec}}\email{}
\author[2]{\fnm{Chandan} \sur{Kumar}}\email{}
\author[3]{\fnm{Emanuele} \sur{Marsili}}\email{}

\author[2]{\fnm{Fengmin} \sur{Du}}\email{}
\author[5]{\fnm{Daniel} \sur{Weber}}\email{}

\author[4]{\fnm{Jasper Simon} \sur{Krauser}}\email{}
\author[2]{\fnm{Elvira} \sur{Shishenina}}\email{}
\author[1]{\fnm{David} \sur{Mu\~noz Ramo}}\email{}

\affil[1]{\orgname{Quantinuum}, \orgaddress{\street{Terrington House, 13-15 Hills Road}, \city{Cambridge} \postcode{CB2 1NL}, \country{United Kingdom}}}

\affil[2]{\orgname{BMW Group}, \orgaddress{\city{Munich} \postcode{80788},  \country{Germany}}}

\affil[3]{\orgname{Airbus, Central Research \& Technology}, \orgaddress{\street{Pegasus House Aerospace Ave}, \city{Bristol} \postcode{BS34 7PA},  \country{United Kingdom}}}

\affil[4]{\orgname{Airbus, Central Research \& Technology}, \orgaddress{\street{Willy-Messerschmidt-Str. 1}, \city{Taufkirchen} \postcode{82024}, \country{Germany}}}

\affil[5]{\orgname{Aerostack GmbH}, \orgaddress{\city{Dettingen an der Erms} \postcode{72581}, \country{Germany}}}


\abstract{Hydrogen has emerged as a promising energy source, holding the key to achieve low-carbon and sustainable mobility. However, its applications are still limited by modest conversion efficiency in the electrocatalytic oxygen reduction reaction (ORR) within fuel cells. Consequently, the development of novel catalysts and a profound understanding of the underlying reactions have become of paramount importance. The complex nature of the ORR potential energy landscape and the presence of strong electronic correlations present challenges to atomistic modelling using classical computers. This scenario opens new avenues for the implementation of novel quantum computing workflows to address these molecular systems. Here, we present a pioneering study that combines classical and quantum computational approaches to investigate the ORR on pure platinum and platinum/cobalt surfaces. Our research demonstrates, for the first time, the feasibility of implementing this workflow on the H1-series trapped-ion quantum computer and identify the challenges of the quantum chemistry modelling of this reaction. The results highlight the involvement of strongly correlated species in the cobalt-containing catalyst, suggesting their potential as ideal candidates for showcasing quantum advantage in future applications.}


\maketitle

\subsection*{Introduction} \label{sec1}
\begin{figure}[h]%
\centering
\includegraphics[clip, width=1.0\textwidth]{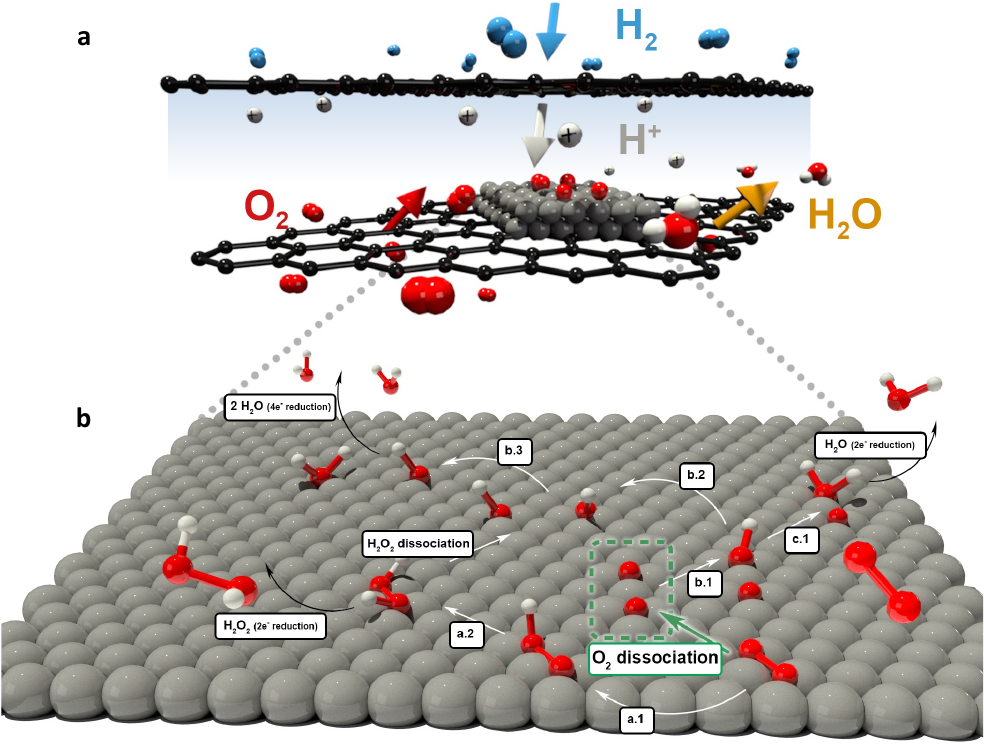}
\caption{\textbf{PEMFC schematic and complexity of the multistep catalytic process promoted by platinum.} \textbf{a}, Schematic of a PEMFC with anodic (oxidation): $\mathrm{H_2 \rightarrow 2H^+ + 2e^-}$ and cathodic (reduction): $\mathrm{\frac12\,O_2 + 2H^+ + 2e^- \rightarrow H_2O}$ reactions. \textbf{b}, The ORR mechanism involves a multistep path that incorporates both dissociative and proton-assisted steps, ultimately leading to the formation of \ce{H2O} or \ce{H2O2}. The reduction of molecular oxygen leads to the formation of atomic oxygen (framed in green), hydroxyl radicals (reactions b.1 and b.2), hydroperoxy radicals (reaction a.1), and ultimately, with the release of \ce{H2O2} (reaction a.2) or \ce{H2O} (reactions b.3 and c.1)~\cite{zhang2007}. The dissociation of \ce{O2}, highlighted in green, is identified as the rate-determining step in the overall ORR mechanism. The specifics of this step will be examined using pure platinum and platinum/cobalt catalysts.}\label{fig:group1}
\end{figure}

Decarbonization represents one of the most pressing yet ambitious goals humanity faces in the coming decades to reduce \ce{CO2} and greenhouse gas emissions. In this context, the pursuit of innovative green energy sources takes precedence. Currently, the mobility sector mostly relies on hydrocarbons due to their availability and engines capable of an efficient conversion from chemical energy into mechanical work. In recent times, hydrogen has emerged as a viable alternative to hydrocarbon fuels thanks to its advantageous energy-to-weight ratio, making it potentially competitive to battery-powered electric vehicles. 
Hydrogen can be either utilized in combustion engines, in a manner akin to the combustion of hydrocarbons, or its chemical energy can be transformed into electricity through a fuel cell which then powers an electric motor. The latter circumvents the limitations of the Carnot cycle and achieves a more efficient energy conversion process~\cite{hosseini2020an}.
Among various fuel cell architecture, the proton-exchange membrane fuel cell (PEMFC), illustrated in Fig.~\ref{fig:group1}a, is the most developed and well understood fuel cell technology compared to solid oxide or alkaline fuel cells. In a PEMFC the molecular hydrogen is oxidized at the anode, resulting in the production of protons (\ce{H+}). The \ce{H+} ions then migrate through the proton-exchange membrane towards the cathode, where oxygen is reduced on a catalyst. The interaction of the \ce{H+} ions with the reduced oxygen atoms leads to the formation of \ce{H2O}, which is the main product of the hydrogen fuel cells.
Despite its conceptual simplicity, the development of highly efficient fuel cells -- critical for a more sustainable application in the mobility sector -- remains an unresolved challenge. One significant bottleneck lies in the ORR, which experiences a substantial kinetic overpotential. In an ideal reversible hydrogen electrode (RHE), the combined net transfer of four protons and electrons to molecular oxygen generates a potential of $1.23$ V$_\text{RHE}$ in acidic media. However, in current fuel cells, the observed voltage is less than 0.9 V at any usable current output~\cite{Keith2010}.

The exact cause of this overpotential remains undisclosed to date~\cite{chen18, norskov04} despite extensive research has been already dedicated to study the ORR mechanism. Due to the complex nature of the reactions involved~\cite{zhang2007} (see Fig.~\ref{fig:group1}b), providing an explicit description of the entire reaction dynamics proves to be impractical. By analyzing each reaction step separately, we can extract valuable information into both reaction kinetics -- related to its activation barrier -- and thermodynamic stability -- linked to the energy difference between products and reactants. This approach was applied to investigate the multistep evolution of ORR on pure platinum catalysts, identifying two rate-determining steps in the entire reaction network~\cite{Keith2010}: i) the proton-coupled process \ce{O_2^* + H^+ + e^- ->[Pt] HO_2^*} (Fig.~\ref{fig:group1}b, step a.1)~\cite{Damjanovic1967, Sepa1987} and ii) the reductive and dissociative adsorption \ce{O_2 ->[Pt] 2O} (Fig.~\ref{fig:group1}b, green arrow)~\cite{Clouser1993, yeager1993} mediated by an electron exchange. 

The catalyst plays a pivotal role in shaping the reaction dynamics and, despite various metals have demonstrated potential applicability in catalyzing the ORR, to date, platinum and its combination with cobalt or nickel~\cite{asara2016, shao2016, kulkarni2018, stamenkovic2007, zou2015, wang2013} are considered among the most promising. These catalysts exhibit favorable structural and electronic properties that effectively destabilize intermediates, thereby increasing the kinetic rate by reducing activation energy barriers~\cite{zhang2021, molmen2021, wang2021}. Hence, modelling the electronic interaction between these intermediates and the catalyst surface is crucial to understand the mechanism of the ORR and establish the basis for developing more efficient fuel cells.

Computational chemistry stands out as the perfect tool for characterizing the structural details of the ORR and quantifying the kinetic and thermodynamic factors associated with the rate-determining reaction steps. Currently, density functional theory (DFT) has been the predominant method for studying the ORR, largely owing to its favorable scaling with the system size. However, a notable drawback is the reliance of DFT accuracy on the choice of an approximate functional, which often result in a poor description of strongly correlated systems~\cite{cohen2012}. More accurate wave-function-based techniques, including coupled cluster singles and doubles (CCSD), full configuration-interaction quantum Monte Carlo (FCIQMC), and density matrix renormalization group (DMRG), are instead more reliable for strongly-correlated systems~\cite{gruber2018a, gruber2018b, Tsatsoulis2018, booth2013, DMRG_rev_Reiher2020}. However, the computational resource demands of these methods escalate significantly with system size. This limitation renders these techniques impractical for the majority of real-world applications in catalysis and materials science, including the ORR, due to their computational requirements.

In this context, quantum computing emerges as an attractive candidate to overcome the poor scaling of conventional quantum chemistry. Quantum computing utilizes inherently quantum objects to efficiently represent electronic wavefunctions. This approach enables the allocation of fewer resources and a significant reduction in the number of required logic operations.

This paper introduces a completely novel hybrid workflow that combines the strengths of both quantum and classical computing techniques to model the ORR. Specifically, we explore how the kinetic and thermodynamic aspects of the reductive and dissociative adsorption reaction (\ce{O_2 -> 2O}), occurring at the surface of platinum (Pt) and platinum-capped cobalt (hereafter referred to as Pt/Co) catalysts, can be quantified. We use an embedding method that couples the relevant electronic degrees of freedom of a representative subsystem (fragment), described using the variational quantum eigensolver (VQE)~\cite{Peruzzo2014, tilly2022}, with an environment treated within the mean-field theory and the N-electron Valence State Perturbation Theory (NEVPT2) approach to account for the electronic dynamical correlation. 

Our study demonstrates, for the first time, that quantum computing is already mature enough to prototype a realistic application. More importantly, we believe that it will offer a significant advantage in catalysis, especially in scenarios involving strongly correlated electrons, as exemplified in the case of Pt/Co, even more, as hardware continues to advance towards fault-tolerant machines capable of supporting highly demanding quantum algorithms like the quantum phase estimation (QPE) approach~\cite{lee2021}.

\subsection*{Embedding scheme and energy evaluation via quantum computing}

The objective of the embedding scheme is to couple a few electronic degrees of freedom, potentially strongly correlated, with the electronic degrees of freedom of the rest of the system, referred to as environment or bath. This integration capitalizes on the localized nature of interactions between the adsorbate and the catalyst surface at the landing site. In this way, we can utilize correlated methods, \textit{i.e.}, post-Hartree-Fock techniques, to model the interactions at the adsorption site while, for the rest of the system, we employ efficient mean-field theories like HF or DFT~\cite{Leighton2020, Nusspickel2022}.

\begin{figure}
    \centering
    \includegraphics[width=\textwidth]{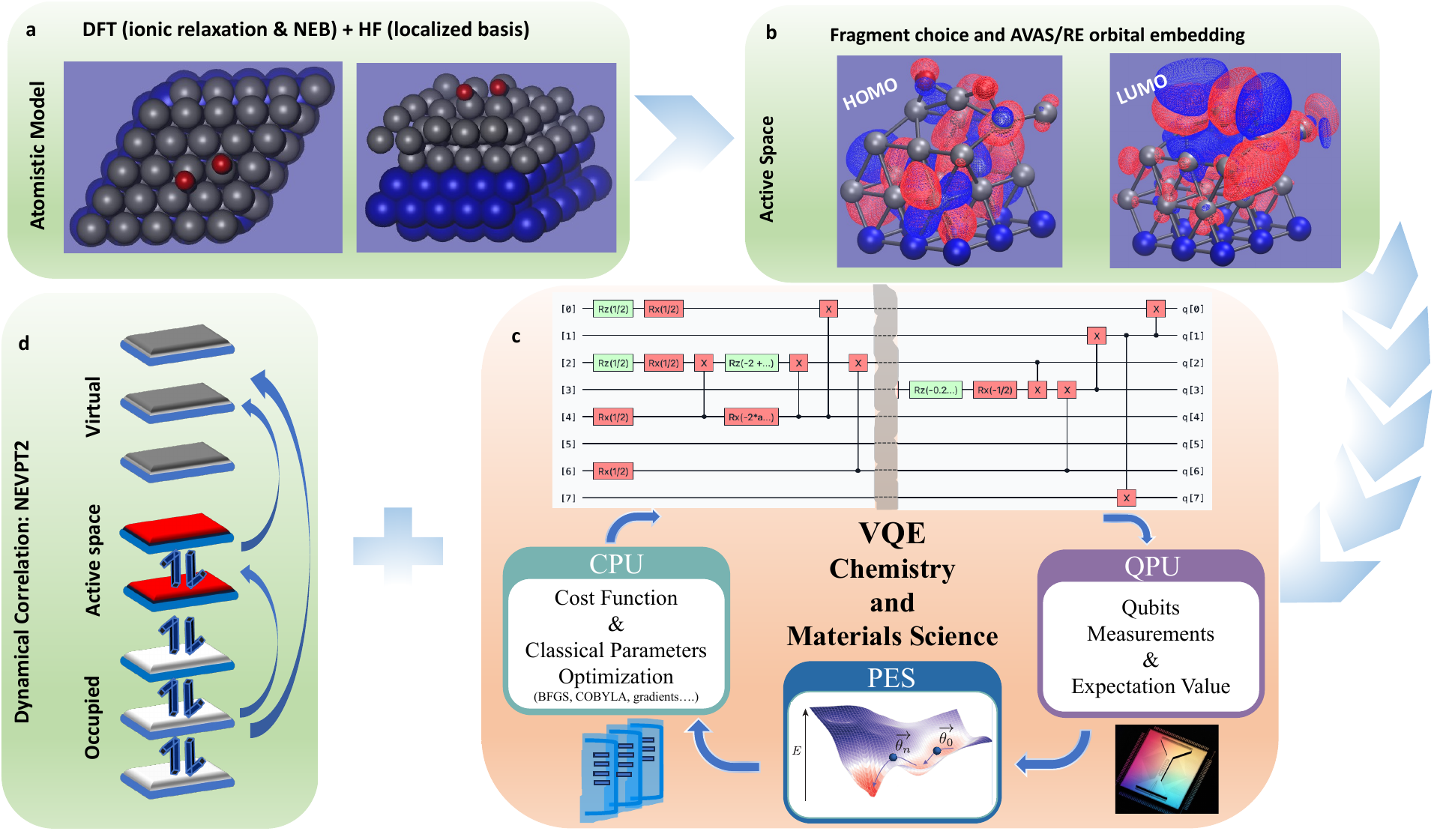}
    \caption{\textbf{The hybrid quantum-classical workflow}. \textbf{a,} An atomistic model of the whole system is generated, firstly geometrically relaxed using a DFT approach and, then,  treated at the HF level of theory with a localized basis. \textbf{b,} An atomistic representation of the (electron correlated) embedded fragment is carefully chosen and an AVAS/RE procedure is performed to find the best embedded orbital set for the active space. \textbf{c,} Energy calculation for the embedded active space including static correlation via a hybrid quntum-classical ADPT-VQE procedure is performed by optimizing the variational parameters for the chosen ansatz and active space (CPU: Intel Xeon; quantum device QPU: Quantinuum H1-series hardware/noisy emulator). \textbf{d}, To estimate the contribution of the dynamical correlation to the total energy, a classical post-processing NEVPT2 calculation is made using the set of RDMS, up to the $4^{th}$ particle order, measured after the optimization cycle in step \textbf{c} is completed.}
    \label{fig:group2}
\end{figure}

A crucial step in this process is partitioning the electronic degrees of freedom, which we accomplish through the automatic valence active space/regional embedding (AVAS/RE)~\cite{AVAS} approach (see Fig.~\ref{fig:group2}b). This method generates highly localized molecular or crystal orbitals, selecting the most strongly correlated and chemically relevant while excluding those mainly connected to the bath (see the Method section for more details).

Given the limited capabilities of modern quantum hardware, we restrict the active space to 2 electrons in 3 orbitals (2e,3o), necessitating 6 qubits, and to (4e,4o), requiring 8 qubits, for pure platinum and Pt/Co, respectively. In both cases, we configure the electrons in the lower-energy spin-orbitals, establishing the initial Fock state that serves as the Hartree-Fock (HF) reference point. The circuit diagram for the Pt/Co experiment is depicted in Fig.~\ref{fig:group2}c.

After initializing the Fock state, we construct the wavefunction ansatz for measuring the expectation value of the electronic Hamiltonian. A practical approach, well-suited for the current noisy quantum computers, involves employing the VQE~\cite{Peruzzo2014, tilly2022}. In this method, a parameterized wavefunction is prepared using a specific ansatz while the underlying parameters are optimized on a classical computer to minimize the expectation value of the electronic Hamiltonian (see Fig.~\ref{fig:group2}c). Our parametrized wavefunction corresponds to the generalized unitary coupled cluster (k=1)-UpCCGSD~\cite{Lee2019}. To reduce the wavefunction degrees of freedom even more, we further simplify the ansatz by applying the Fermionic VQE-ADAPT algorithm~\cite{Grimsley2019}. This algorithm progressively constructs the ansatz by sequentially incorporating operators that contribute most to lowering the VQE energy, guaranteeing a moderate circuit depth as well as a drastic reduction in the number of parameters to optimize. Finally, prior to the measurement, we employed the partition measurement symmetry verification (PMSV) error mitigation procedure~\cite{Yamamoto2022} to suppress spurious symmetry-breaking errors in the particle number and spin parity conditions.

After the state preparation, we measure the expectation value of the electronic Hamiltonian and the elements of the reduced density matrices (RDMs). To optimize the measurement step, we divide the overall Pauli words into mutually commuting sets in which, each set defines a measurement circuit. In our case, we needed a batch as large as 12 and 19 circuits for the energy, for pure Pt and Pt/Co, respectively. The knowledge of the RDMs allows us to improve the VQE energy by constructing, on a classical computer, the second order perturbation correction terms (in the specific the NEVPT2 approach, see scheme Fig.~\ref{fig:group2}d). These perturbation terms recover the remaining dynamical correlation not captured by the VQE wavefunction~\cite{krompiec2022}.

\section*{The \ce{O2} dissociation on the catalytic surface of pure platinum} 

\begin{figure}[h!]%
    \centering
    \includegraphics[width=\textwidth]{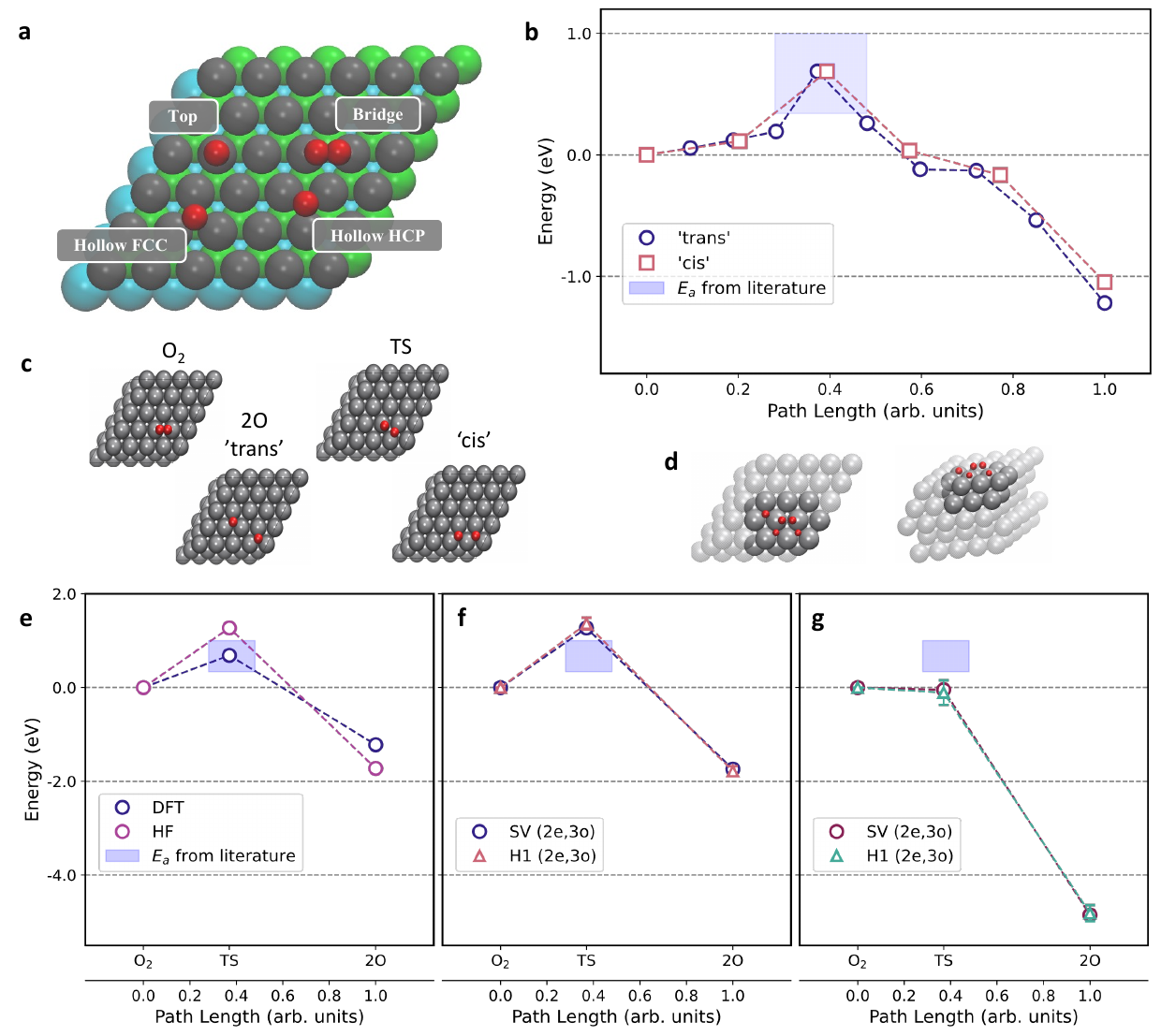}
    \caption{\textbf{NEB, DFT, HF, VQE and quantum experiments for oxygen dissociation on pure platinum surface.} \textbf{a}, Structure of the ABC FCC(111) layer composition. A-layer: surface layer (grey), B-layer: first sub-layer (green), C-layer: second sub-layer (cyano). Adsorption sites: T=top; B=bridge; HCP=hollow hexagonal close-packed, facing B-layer Pt atom;
    FCC=hollow FCC, facing C-layer Pt atom. \textbf{b}, NEB profiles connecting the reactant with the 'cis' (purple circles) and 'trans' (violet squares) products for the \ce{O_2 ->[Pt] 2O} reaction. 
    \textbf{c}, Top view of the atomistic platinum slabs with oxygen reactant ($O_2$), transition state (TS) and dissociated adsorbates ('trans' and 'cis'). \textbf{d}, top and lateral view of the pure platinum model with the 'Pt19O2' fragment in dark gray. All the possible positions for adsorbed molecular and atomic oxygen in red spheres, also associated with 'cis' and 'trans' products. \textbf{e}, \textbf{f}, \textbf{g}, Energy profiles along the NEB reaction path: \textbf{e}, the mean field calculations are DFT (in purple) and HF (in violet); \textbf{f}, energies computed using our quantum computing algorithm based on VQE for the SV (in violet) and the resulting measurement from the H1 quantum hardware (in light-red); \textbf{g}, energies computed using our quantum computing algorithm based on VQE and corrected by including dynamic correlation (VQE+NEVPT2 energies) for the SV (in violet) and the resulting measurement from the H1 quantum hardware (in green). The error bars are calculated as a standard deviation around the mean value obtained by bootstrapping 10k shots in 10 batches each for Quantinuum H1 hardware device measurements. Activation energy $E_a$ available from literature~\cite{Montemore2018} depicted as light-blue area.}
    \label{fig:group3}
\end{figure}

Identifying the adsorption sites on the catalyst surface is a key step in modeling the \ce{O_2 ->[Pt] 2O} reaction occurring on the surface of Pt. Concerning the \ce{O2} molecule, it can adsorb in a bridge position between two nearest-neighbor platinum atoms. In contrast, the oxygen atoms -- \textit{i.e.}, the products -- can occupy a hollow hexagonal close-packed (HCP), hollow face-centered cubic (FCC), or top position~\cite{Xue2018, Pushpa2003, Crljen2003}, as illustrated in Figure~\ref{fig:group3}a.

Our investigation begins with an analysis of the \ce{O_2 ->[Pt] 2O} reaction's thermodynamics, focusing on the energy difference between reactants and products. As expected, the \ce{O_2} molecule adsorbs onto the Pt surface exclusively in a 'bridge' configuration (see Figure~\ref{fig:group3}c), with associated electronic energy $E_{O_2}$, taken as reference. Conversely, our DFT atomistic model of the Pt slab reveals two possible arrangements for the atomic oxygen in hollow FCC~\cite{Xue2018}, from now on termed 'cis' and 'trans' (also Fig.~\ref{fig:group3}c). These two products possess respective electronic energies $E^{\text{cis}}$ and $E^{\text{trans}}$ and demonstrate a notable DFT-calculated energy difference of $E^{\text{trans}}-E^{\text{cis}} = -0.17$ eV between them. The corresponding \ce{O_2} dissociation energy ($E_d$) is then computed as $E_d^{\text{cis}} = E^{\text{cis}} - E_{O_2}$ and $E_d^{\text{trans}} = E^{\text{trans}} - E_{O_2}$ for the formation of the 'cis' and the 'trans' products, respectively (see Fig.~\ref{fig:group3}b).

The reaction kinetics is easily estimated by the Arrhenius equation, where the activation energy ($E_a$) figures as a key component. The calculation of the $E_a$ can be reduced to the estimation of the electronic energy difference between the reactants and the transition state (TS). The TS is characterized by the two oxygen atoms positioned in between a hollow and a top position (see Fig.~\ref{fig:group3}b and TS structure in Fig.~\ref{fig:group3}c). Pinpointing the exact TS geometry is often a formidable challenge; thus, we applied the nudged elastic band method (NEB)~\cite{jonsson1998}, as implemented in Quantum Espresso (QE)~\cite{QE_09, QE_17}, to approximate the geometry of the TS. Despite the two products being identified, our findings reveal identical transition states for the reaction paths resulting in a unique $E_a$ value of 0.69 eV, closely aligning with the range of literature-reported $E_a = 0.67 \pm 0.33$ eV~\cite{Montemore2018}. The NEB calculations also delineate the minimum energy paths (MEPs) which are illustrated for the 'cis' and 'trans' pathways in purple squares and violet circles, respectively, in Fig.~\ref{fig:group3}b.

To select the best AVAS/RE active space, we first selected the best fragment of 'Pt19O2' atoms from consideration on the electronic density difference as described in Supplementary Fig. S1. We then studied the convergence of the dissociation energies ($E_d$) for both ‘cis’ and ‘trans’, and their relative stability ($E^{trans}-E^{cis}$), with respect to the AVAS
overlap threshold (see description in Supplementary Fig. S2a). We also tested energy convergence against the number of layers in the surface model by choosing a narrower AVAS threshold window around 0.2 (see Supplementary Fig. S2b). We found that a 5-layer slab (as described in the caption Fig.~\ref{fig:group3} and depicted in its panel c) with an AVAS threshold=0.18 was the best possible setup.

Since for the following hybrid quantum-classical workflow we take advantage of HF orbital localisation as implemented in the PySCF package~\cite{sun2020,sun2018}, in the panel of Fig.~\ref{fig:group3}e we compare how HF compares with DFT energy profiles, shown in violet and purple, respectively. By comparing the energies for the TS point, we highlight that HF predicts a higher energy barrier than DFT. The same is also observed in the VQE calculations (Fig.~\ref{fig:group3}f), where both the state vector (SV) and the H1 quantum hardware calculations are, surprisingly, very close to the HF energies, and they significantly overestimate the energy barrier predicted with DFT. From this comparison, we can draw three conclusions. 1) The data generated by the H1 quantum hardware demonstrates high quality, evidenced by minimal discrepancies from the state vector, with energy differences of $\Delta (E_{TS}-E_{SV}) =\, 0.300$eV for activation energy and $\Delta (E^{\text{trans}}-E_{SV}) =\, 0.163$eV for dissociation energy (full set of values also for H1-E noisy emulator in Supplementary Tab. S2). This is particularly promising as it indicates that, despite the hardware still being in its early stages of development, the workflow employed here enables us to effectively address realistic models without being excessively constrained by noise limitations. 2) The overall deviation from the DFT reference may originate from discrepancies in the optimal geometries needed by each theoretical level. While expected, this poses a future challenge in optimizing large atomistic models with methods costlier than DFT. 3) The resemblance of the mean-field outcomes to the VQE profiles indicates that static correlation, associated with the selection and dimension of the active space, has limited influence in the case of pure Pt catalyst. In simpler terms, inclusion of a few orbitals in the correlated subspace within the embedding procedure does not greatly improve the results which are, from our prospective, somewhat unexpected.  

Let's now examine the impact of correcting the VQE energy with PT2 (alias, NEVPT2), thus recovering the dynamic electronic correlation. As depicted in Fig.~\ref{fig:group3}g, the addition of the PT2 greatly influences the dissociation and the activation energy, stabilizing both the products and the TS relative to the reactant. At this point a predicted active energy $E_a$ is close to zero, which is not unexpected, once again due to differences between DFT and HF potential energy surfaces. It is also worth noting that: 1) statistical errors evaluated on the set of quantum measurements (as described in caption Fig.~\ref{fig:group3} and listed in Supplementary Tab. S1) are found not larger than $\epsilon = 0.2\, eV$, for $E_a$ in the VQE+NEVPT2 simulation; 2) preliminary tests on the H1-E noisy emulator were also performed and the results, aligned with the actual quantum hardware outcomes, presented in Fig. S2 of the Supplementary Information.

Finally, from this analysis we conclude that, for the \ce{O_2 -> 2O} catalyzed by pure Pt, dynamical correlation emerges as a crucial contribution to the total energy of the system, while static correlation proves to be nearly negligible.

\subsection*{The Oxygen dissociation on platinum and cobalt catalyst} \label{sec3}

In order to simulate the oxygen reduction assisted by the Pt/Co surface, we started with optimising the Co slabs capped with 1, 2 and 3 atomic layers of Pt, denoted 1L, 2L and 3L hereafter (see models in Supplementary Fig. S4 and supercell in Fig.~\ref{fig:group4}b). We observed that the distribution of magnetic moments along the slabs showed a crucial dependency on the thickness of Pt capping, with the magnetism of the surface Pt atoms dramatically decaying with the distance from Co bulk, as shown in the Supplementary Tab. S5.

Subsequently, we proceeded to the calculation of O$_2$ molecule and two separate oxygen atoms bound to the surface layer. Moreover, a frozen surface would lead to a scenario where the computed energies of the P states are approximately $0.1$ eV higher than the R states for all the Pt-layer thicknesses considered. Consequently, it was of paramount importance to include surface relaxation on the Pt/Co model. In this way, we were able to restore the correct energy ordering. For further analysis of ORR on Co coated platinum model, we have chosen to bring forward the 2L system, since: 1) there are experimental indications that it would be the best catalyst~\cite{osti1820884} and 2) because it represents an intermediate case between the 1L model with a lower catalytic effect and the 3L one which is closer to the bulk Pt limit. To calculate the activation energy $E_a$ (see Fig.~\ref{fig:group4}a), we needed to determine the structure of the TS, $\mathrm O_2^*$ on Pt(111) and Pt/Co(111). To this end, we performed a Nudged Elastic Band calculation \cite{jonsson1998}, as implemented in QE~\cite{QE_09, QE_17}. Due to significant reconstruction of the platinum surface layer observed during the geometry relaxation on top of Co(111) core, we performed $\mathrm{O_2 \rightarrow 2O}$ NEB optimisation by letting the 25 Pt atoms adapt to the oxygen molecule dissociation just along the $z$ direction, since the in-plane atomic displacements are negligible due to highly packed $(111)$ arrangement. 

We refined the NEB profile several times, by zooming in on the part of the path containing the activation energy barrier related to the dissociation process and by discarding the diffusion parts, where the energy remained essentially flat along the reaction coordinate and so irrelevant for the determination of $E_a$. In this case, the highly computationally challenging optimisation search was performed by using 10 NEB images and 200 NEB steps. Within DFT, we found a final barrier $E_a$ of about $0.46$\,eV, while the driving force was $E_d=-1.30$\,eV, as shown in Figure~\ref{fig:group4}a. As a result of the complexity of the process, the MEP displayed a structured profile with two potential peaks of almost equal heights. For the rest of this study, we will consider the structure at path length$\sim$0.6 (NEB image number 6) as representative model for the transition state (TS, Figure~\ref{fig:group4}a). 
\begin{figure}
    \centering
    \includegraphics[width=\textwidth]{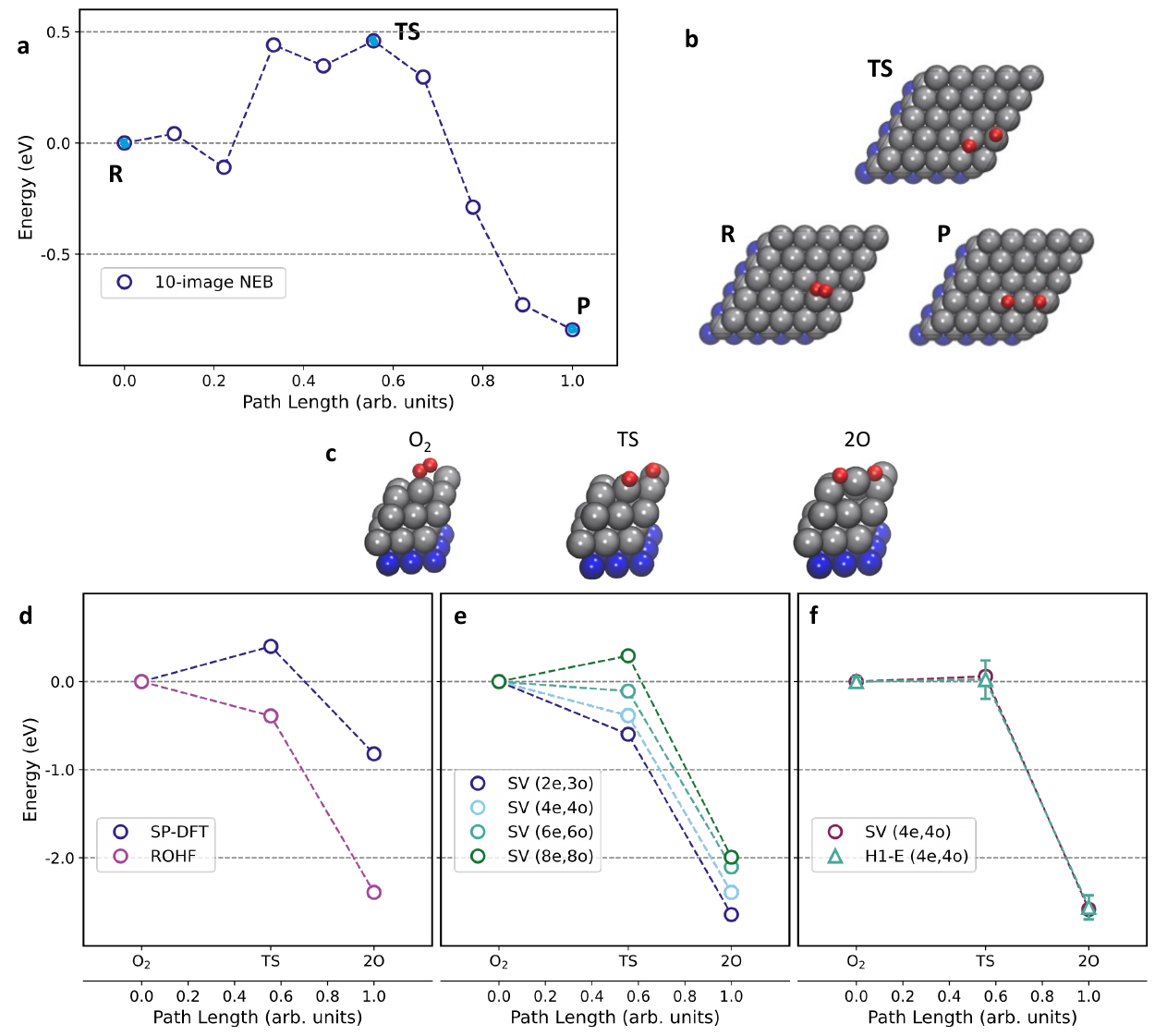}
    \caption{\textbf{NEB, SP-DFT, ROHF, VQE and quantum simulations on H1-E noisy emulator for oxygen dissociation on cobalt and platinum core/shell surface}. 
    \textbf{a}, NEB profile for the oxygen dissociation on Pt/Co catalyst.
    \textbf{b}, top view of the atomistic models depicting positional arrangements of oxygen atoms on top of the cobalt and platinum core/shell surface slab. \textbf{c}, atomistic models of the best molecular fragment of the structures depicted in panel \textbf{c}, used for active space simulations.
    Comparison of: \textbf{d}, restricted open-shell ROHF and spin-polarised SP-DFT energies, \textbf{e}, VQE (tagged as VQE$_1$) energies at different active space size, and \textbf{f}, VQE (tagged as VQE$_2$) energies with PT2 correction on SV and H1-E quantum processor emulator. VQE$_1$ energies (panel \textbf{e}) are based on 4 different active spaces chosen with AVAS. VQE$_2$ refines the (4e,4o) active space used in VQE$_1$.}
    \label{fig:group4}
\end{figure}
The AVAS/RE workflow was organised as follows. A smaller finite fragment was cut out of the optimised periodic structures (full periodic model and molecular fragments depicted in Figs.~\ref{fig:group4}b and \ref{fig:group4}c, respectively). This choice of a smaller fragment was dictated by several considerations: i) the charge difference between R and P states is localised in the area containing the two oxygen atoms and in a 3$\times$3 square of surface Pt atoms, ii) the second Pt layer and at least one Co layer should be included into the small cluster in order to see the effect of magnetism on the oxygen dissociation reaction, as shown in Supplementary Tab. S5, and iii) the 3$\times$3 in-plane fragment was large enough to reproduce the energetics of the PBC (5$\times$5$\times$5) slab. The resulting model comprised $29$ atoms: $2$ oxygen, $18$ platinum and $9$ cobalt atoms. The magnetisation of each Co atom was essentially the same as in the QE calculations for the full periodic model: $\sim 2\mu_B$, so that the whole fragment could be easily treated within PySCF~\cite{sun2018, sun2020} ROHF calculations as a first step. 
In the AVAS procedure the surface Pt $5d$, O $2p$ and $3p$ orbitals were employed as local atomic projectors, even though in an ROHF open-shell situation with finite magnetisation this was not a trivial task, the reason being that half-filled Co $3d$ orbitals, the main source of the total magnetism in the system, contributed to populate the $d$-band manifold near the Fermi level. In our specific case, their correlation contribution had to be removed from the active space, since we were interested in correlations associated with the oxygen dissociation reaction and not with the Co magnetism. ROHF energies and comparison with DFT calculations in Fig.~\ref{fig:group4}d.

We explored several CAS configurations with the single AVAS procedure followed by the state vector VQE~\cite{Grimsley2019}: $(2e,3o)$, $(4e,5o)$, $(6e,6o)$, $(8e,7o)$ and, $(8e,8o)$ respectively. We also checked that all the AVAS active orbitals were always localised on the two oxygen atoms and the surface platinum atoms nearest to the oxygen. The results of this single-shot AVAS procedure are illustrated in Fig.~\ref{fig:group4}e as ``single AVAS VQE'' or VQE$_1$ (where Hartree-Fock results are added as a ``zero-approximation" VQE) and in the Supplementary Material Fig. S5. 
While all the R, TS and P states' energies decrease in accordance with the variational principle, in contrast, their relative stability changes dramatically. Starting from $(8e,7o)$, the activation energy became positive and the driving force increased. They reached the best values with the largest active space in the present work ($(8e,8o)$ adaptive VQE) of $0.293$\,eV for the activation energy and $-1.994$\,eV for the driving force. Compared to QE DFT calculations, the barrier lowered by almost a factor of two, while the driving force increased again by almost a factor of two. The inversion of the trend for the activation energy, while increasing the active space size, suggests the presence of a strong static correlation in the system. From our analysis, it is also clear that most of this correlation could be captured by considering an active space with more fractionally occupied than virtual orbitals. We used these VQE$_1$ calculations as a benchmark for our quantum experiments, indicating that a large active space, prohibitive for modern NISQ devices, is needed to recover the correct energy level ordering.

Similarly to the pure Pt case, in the case of Pt/Co(111), further improvement on the active space treatment, which was beneficial also for the energies, was obtained by using a two-step AVAS/RE procedure. Namely, the CAS active space of $(20e,40o)$ was used at the first step. Subsequently, this effective model was further reduced to a smaller one with a hardware compatible CAS space.
We performed the quantum computations on the NISQ compatible smaller $(4e,4o)$ active space by using the Quantinuum H1-E noisy emulator. This active space required $8$ qubit circuits. In particular, we applied the adaptive VQE, followed by NEVPT2, as described in the Methods section. The comparison with the state vector VQE+QRDM\_NEVPT2~\cite{krompiec2022} and the same active space is also presented in Figure~\ref{fig:group4}f.
In this section, we have performed a series of $5$ runs of  $10k$ shots, each with a calculated standard deviation between $0.02$ eV (for the R state) and $0.1$ eV (for the P state).
We notice that our H1 simulations agree very well with the state vector benchmark for all R, TS and P states, confirming that we were able to efficiently control errors on H1 quantum computer.

\subsection*{Outlook} \label{outlook}

Our study pioneers quantum computing's potential role in simulating hydrogen fuel cell. Our study describes the dissociation of oxygen on catalysts commonly used in fuel cells -- pure platinum and a few surface layers of platinum deposited on a cobalt substrate. Interestingly, we revealed distinct features in the ground state wavefunctions between the two different catalysts. In pure platinum, the wavefunction describing the oxygen dissociation reaction path seems to be dominated by dynamical correlation, a contribution efficiently accounted for in classical computers through perturbation theory approaches. In this case, the system described with quantum computers in our embedding scheme is limited in size. 

When cobalt is added to the catalyst, we observe a significant dependence on the selected active space, where accurate results are achievable only with a large number of explicitly correlated orbitals. This illustrates a typical scenario where the wavefunction grows exponentially with the number of included orbitals, making the problem partially intractable on classical computers but still feasible on quantum machines.

More generally, our approach represents a robust workflow capable of tackling even the most complex catalytic problems, such as the ORR. Thanks to the embedding procedure and the selection of the active space, we ensure the efficient treatment of static correlation energy localized on the adsorption region, maintaining a manageable computational burden on the quantum computer. With access to additional qubits of higher fidelity, expanding the active space provides a straightforward means to improve accuracy by progressively correlating more and more orbitals. Finally, thanks to the perturbation corrections we have an efficient way to recover the remaining dynamical correlation energy whose contribution is often neglected in quantum computing.

Our study lays the foundation for rigorous ab-initio investigations of electrocatalysis models and underscores the potential for early practical applications of quantum computing in fuel cell modeling.

\subsection*{Methods}\label{sec4}

\subsubsection*{First-principles calculations: KS-DFT}\label{sec4_dft}

First-principles simulations for the bulk metal as well as for the multi-layer surface slabs with QE Density Functional Theory (DFT) package~\cite{QE_09, QE_17}. Spin polarised simulations with DFT-D3 van der Waals dispersion correction~\cite{grimme3} for geometry optimisation of structures of interest were performed. In all DFT calculations, we employed the Perdew-Burke-Ernzerhof (PBE) \cite{pbe} GGA exchange-correlation functional together with projector-augmented wave (PAW) pseudopotentials~\cite{Blochl1994}. In the case of Pt surface reaction, the wave-function and charge density cut-offs of 36-47 Ry and 221-448 Ry, respectively, together with a Brillouin zone sampling mesh of (8$\times$8$\times$8) for the bulk system and a Marzari-Vanderbilt smearing of 0.04 Ry were used.

In the Co@Pt case, a wave function cut-off of $60$ Ry with a charge density limit of $450$ Ry were applied by including a sampling k-point mesh of (10x10x1), PAW pseudopotentials and a Marzari-Vanderbilt smearing of $0.01$ Ry.
% The in-house generated PAW pseudopotentials and the plane wave cutoff of $60$ Ry were utilized.

To avoid any density overlapping along non-periodic directions, cubic boxes of at least 10\,\AA\ for small isolated molecules in gas phase such as O$_2$ and $\sim35$\,\AA\, of vacuum on top of multi-layer surfaces ($z$-axis) were adopted.

\subsubsection*{Technical details of the periodic Hartree-Fock calculations}

The periodic HF mean-field calculations have been performed by using the Los Alamos National Laboratory
(LANL2DZ) basis/pseudopotentials of the double-$\zeta$ type both for oxygen and platinum atoms~\cite{li2003} (full 8 and 18 electrons, respectively, treated explicitly and for Pt an effective core potential (ECP) for the remaining electrons treated as core particles), with a 'Fermi-Dirac' smearing as a continuous step function at the Fermi level with a tuning parameter $\sigma = 0.005\; Ha$ and a (6x6x6) k-point mesh to ensure total energy convergence. 

The same setup has been applied to check convergence stability on the (5x5xN with N=1,3,5) platinum model under investigation, by mapping the Brillouin zone at the only $\Gamma$-point and using a Gaussian grid-based approach as the density fitting of choice (less accurate than the Fast Fourier Transform (FFT-based) one but having lower memory requirements) as implemented in PySCF~\cite{sun2018, sun2020}.

% \subsection{Nudged elastic band (NEB)}\label{sec4_neb}

\subsubsection*{AVAS and Regional Embedding}
In order to embed the quantum computations on a smaller system into the full ab-initio system, we have used our own implementation of Regional Embedding (RE)~\cite{RegionalEmbedding}, a variant of the Atomic Valence Active Space (AVAS) method~\cite{AVAS}. Within AVAS, the active space is constructed by first selecting a list of atomic orbitals (projectors), defined as spherically-averaged ground state HF orbitals of free atoms in a minimal basis (MINAO). Subsequently, we have computed an overlap matrix of occupied orbitals projected into the space of these selected atomic orbitals:
\begin{align}
% [S^{A}]_{ij} = \langle i|\hat{P}|j \rangle 
[S^{A}]_{ij} = \langle i | \^{P} | j \rangle
\end{align}
Next, a matrix of eigenvectors $[ U]_{ij}$ is computed such that:
\begin{align}
    S^{A}  U =  U
   \operatorname{diag}(\sigma_1,\ldots,\sigma_{N_\mathrm{occ}})
\end{align}
At this point, there are at most as many non-zero eigenvalues as there are projectors. Moreover, this matrix defines a rotation of the occupied orbitals, which separates them into two groups: those which have non-vanishing overlap with the target atomic orbitals ($\sigma_i\neq 0$) and the remaining ones which have exactly zero overlap with our target space. The latter can stay inactive (as core orbitals), the former are the AVAS active orbitals among the occupied subset. An analogous transformation is performed on the virtual orbitals. To further reduce the size of the active space, the orbitals with the overlap $\sigma_i$ lower than a certain threshold (from now on, here referred to as AVAS Overlap Threshold) are excluded. The Regional Embedding \cite{RegionalEmbedding} method differs from AVAS in the way the virtual orbitals are constructed: instead of the minimal basis, the basis functions of the actual computational basis are used as atomic projectors. 

\subsubsection*{Multireference VQE calculations}
To account for dynamic correlation effects, we have used our proprietary method QRDM\_NEVPT2, the details of which have been recently published by Krompiec et al.~\cite{krompiec2022} This procedure calculates a perturbative correction arising from excitations within the whole RE active space. We have used the VQE-ADAPT ansatz building approach, and all the classical and quantum computing simulation were performed inside InQuanto (https://www.quantinuum.com/products/inquanto) the Quantinuum's Computational Chemistry Platform and TKET via PyTKET compiler~\cite{Sivarajah2021}, using the Qulacs statevector emulator \cite{suzuki2021qulacs} and Quantinuum noisy emulator H1-2E. 

\backmatter

%\bmhead{Supplementary information} Here reference to SI file if needed.

\bmhead{Acknowledgments}
The authors thank Duncan Gowland, Gabriel Greene-Diniz for their feedback on the manuscript. 
The authors also thank Nathaniel Burdick, John Children, Vanya Eccles, Isobel Hooper, Brian Neyenhuis and Jenni Strabley for assistance in the software and hardware experiments. 
The authors performed this work partially using the mat3ra 
platform, a web-based computational ecosystem for the development of new 
materials and chemicals~\cite{Exabyte}. 

% \begin{appendices}

% \section{Supplementary Material} \label{secA1}

% \end{appendices}

%\newpage

%\newpage
%\bibliography{references}% common bib file



\end{document}

\title[Supplementary Information:
Platinum-based Catalysts for ORR
 simulated with a
Quantum Computer]{Supplementary Information: Platinum-based Catalysts for Oxygen
Reduction Reaction simulated with a
Quantum Computer}


\author*[1]{\fnm{Cono} \sur{Di Paola}}\email{cono.dipaola@quantinuum.com}
\author*[1] {\fnm{Evgeny} \sur{Plekhanov}}\email{evgeny.plekhanov@quantinuum.com}
\author[1]{\fnm{Michal} \sur{Krompiec}}\email{}
\author[2]{\fnm{Chandan} \sur{Kumar}}\email{}
\author[3]{\fnm{Emanuele} \sur{Marsili}}\email{}
\author[2]{\fnm{Fengmin} \sur{Du}}\email{}
\author[5]{\fnm{Daniel} \sur{Weber}}\email{}

\author[4]{\fnm{Jasper Simon} \sur{Krauser}}\email{}
\author[2]{\fnm{Elvira} \sur{Shishenina}}\email{}
\author[1]{\fnm{David} \sur{Mu\~noz Ramo}}\email{}

\affil[1]{\orgname{Quantinuum}, \orgaddress{\street{Terrington House, 13-15 Hills Road}, \city{Cambridge} \postcode{CB2 1NL}, \country{United Kingdom}}}

\affil[2]{\orgname{BMW Group}, \orgaddress{\city{Munich} \postcode{80788},  \country{Germany}}}

\affil[3]{\orgname{Airbus, Central Research \& Technology}, \orgaddress{\street{Pegasus House Aerospace Ave}, \city{Bristol} \postcode{BS34 7PA},  \country{United Kingdom}}}

\affil[4]{\orgname{Airbus, Central Research \& Technology}, \orgaddress{\street{Willy-Messerschmidt-Str. 1}, \city{Taufkirchen} \postcode{82024}, \country{Germany}}}

\affil[5]{\orgname{Aerostack GmbH}, \orgaddress{\city{Dettingen an der Erms} \postcode{72581}, \country{Germany}}}

\maketitle

\tableofcontents

\pagebreak

\section{Calculated statistical error on quantum device}
\begin{table}[h]
\begin{center}
\caption{VQE and VQE+PT2 statistical error $\epsilon$, calculated as a standard deviation by bootstrapping 100k and 10k shots in 10 batches each for Quantinuum H1-E noisy emulator and H1 device measurements, respectively. $E_a=E_{TS}-E_{O_2}$ and $E_d=E_{2O}-E_{O_2}$ represent the activation and dissociation energies, respectively. E$_{2O}$ is the total energy for the 'trans' configuration. All the measurements on H1-E and H1 include PMSV error mitigation correction}\label{table:vqe_deviation3}%
\begin{tabular}{@{}lrr|rr@{}}
\toprule
& \multicolumn{2}{c}{$VQE$} & \multicolumn{2}{c}{$VQE+PT2$}\\
\\
 & H1-E  &  H1  & H1-E & H1\\
\midrule
$\epsilon(E_{a})$ [eV] &  0.03993 & 0.12936 & 0.34386  &  0.26283\\
$\epsilon (E_d)$ [eV] & 0.02747  & 0.10834  & 0.03050  &  0.17411 \\
\botrule
\end{tabular}
\end{center}
\end{table}

\pagebreak

\section{Energy difference: experiments vs state vector}
\begin{table}[h]
\begin{center}
\caption{VQE and VQE+PT2 calculated distance $\Delta$ of total averaged energies with respect to StateVector ($E_{SV}$) calculations performed using Qulacs backend in eV units. Total energy calculated as an average by bootstrapping 100k and 10k shots in 10 batches each for Quantinuum H1-E noisy emulator and H1 device measurements, respectively. All the measurements on H1-E and H1 include PMSV error mitigation correction}\label{table:vqe_deviation2}%
\begin{tabular}{@{}lrr|rr@{}}
\toprule
& & \multicolumn{2}{c}{$\Delta (E-E_{SV}) \, \mathrm{[eV]}$} \\
\\
& \multicolumn{2}{c}{$VQE$} & \multicolumn{2}{c}{$VQE+PT2$}\\
\\
Species & H1-E  &  H1  & H1-E & H1\\
\midrule
O$_2$ &  0.14965 &  0.20835 &  0.04404 &  0.02131 \\
TS & 0.17522  & 0.29947 &  0.17557 &  $-$0.03943 \\
2O(`trans') &  0.10064  &  0.16303 &  $-$0.01634 &  0.05915 \\
\botrule
\end{tabular}
\end{center}
\end{table}

 \pagebreak

 \section{Density difference for fragment choice}
 
\begin{figure}[h]%
\centering
\includegraphics[width=0.9\textwidth]{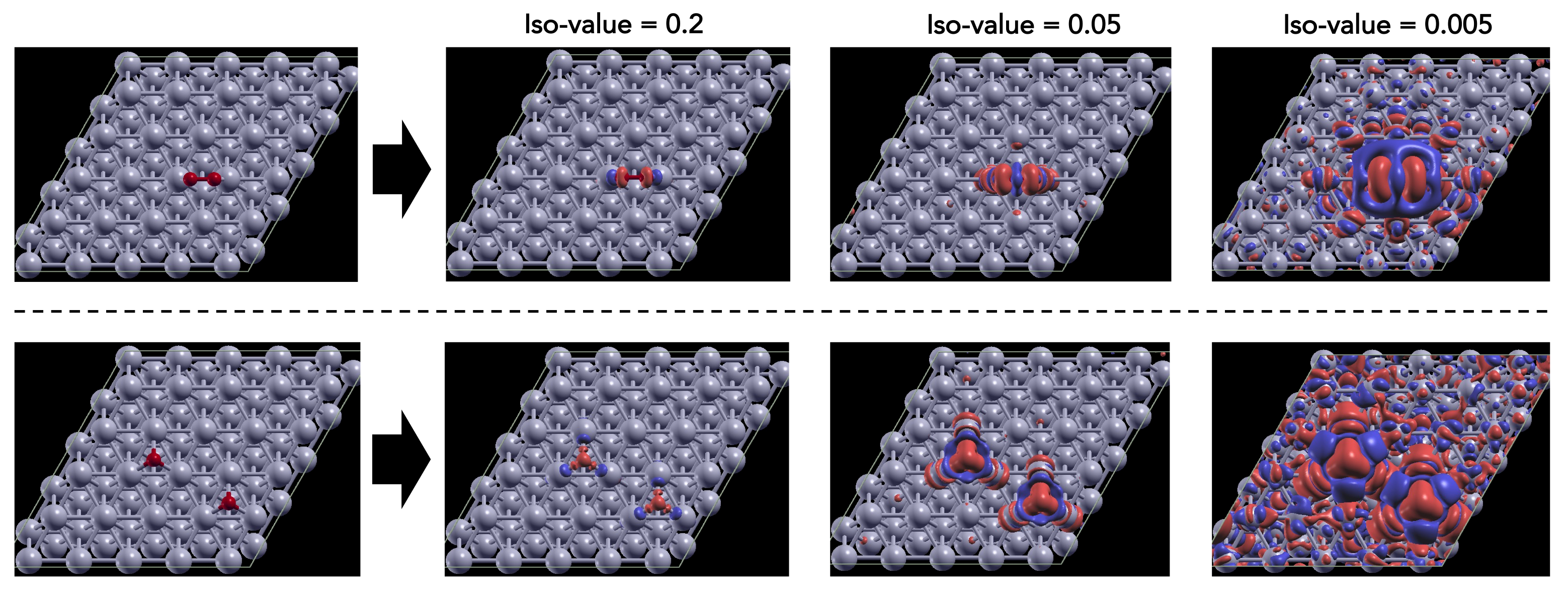}
\caption{DFT density difference: $\Delta \rho = \rho _\mathrm{Pt/OO}-(\rho _\mathrm{Pt}+\rho _\mathrm{OO})$ (where $\rho$ represents the electron density of the single atomic structure), for the two end systems Pt/O$_2$ (upper strip) and Pt/2O `trans' arrangement (lower strip) taken at different iso-values (in units of electron/bohr$^3$). Red and blue volumes represent loss or gain in electron density, respectively, when the adsorbate+substrate is formed from the isolated constituents. It is clear that even at small density iso-values of the order of $10^{-3}$ $e^-$/bohr$^3$ the adsorption/dissociation process of the $O_2$ molecule strictly pertubs the platinum surface electron density mainly at the landing site.}\label{fig:density_diff}
\end{figure}

\pagebreak

\section{Pure platinum surface: AVAS/RE threshold convergence}
 
\begin{figure}[h]%
\centering
\includegraphics[width=\textwidth]{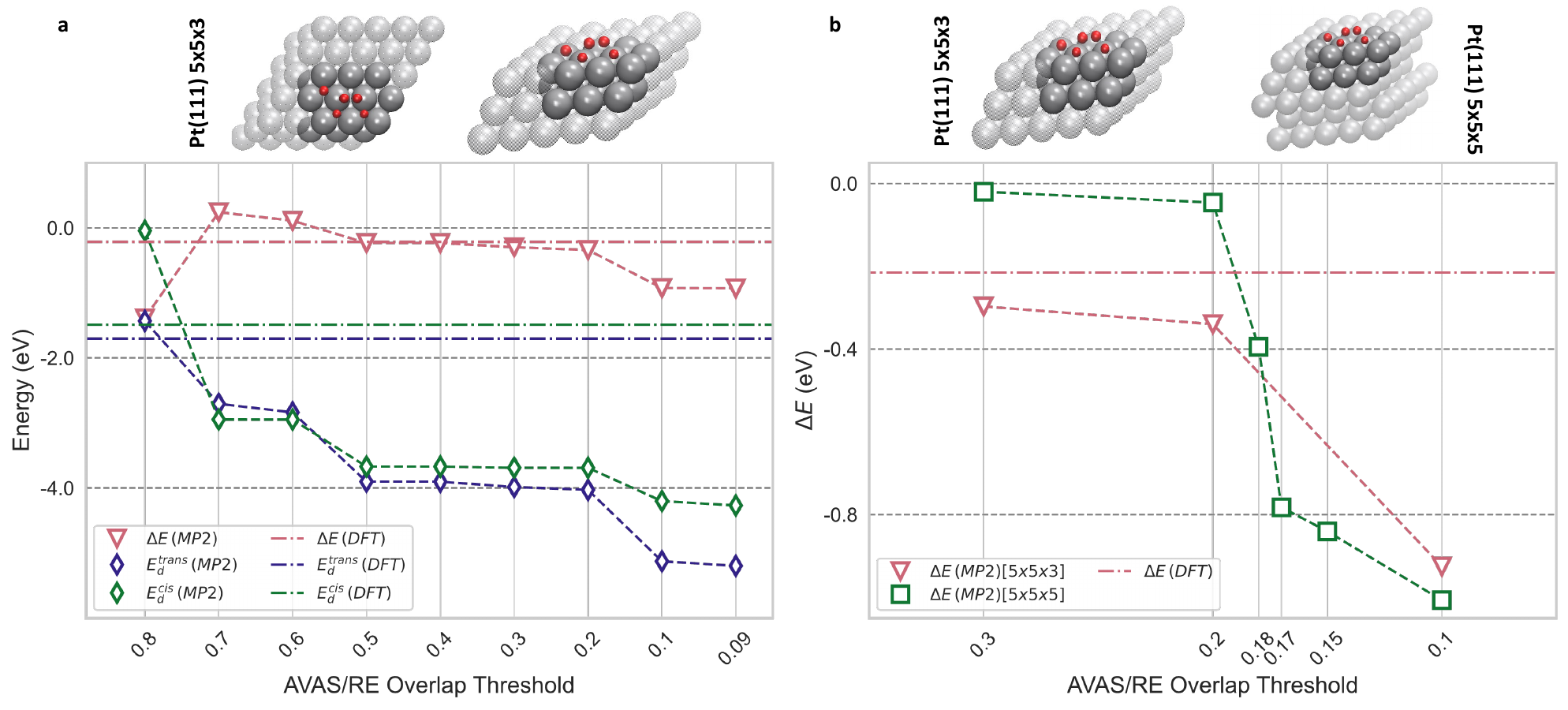}
    \caption{\textbf{Strategy adopted to calculate AVAS/RE threshold.} \textbf{a, upper panel}:  ’Pt19O2’ fragment in a 5x5x3 slab model as solid spheres where the red oxygen atoms show all possible sites considered occupied in this work; \textbf{a, lower panel}: MP2 energies calculated as dissociation energies $E_d$ for ‘trans’ (purple line/diamonds) and ‘cis’ (green line/diamonds) and relative stability of the two isomers $E_{trans}-E_{cis}$ (red line/triangle-down). DFT reference energies are also included. It is clear that the MP2 energy plateau represents the converged AVAS threshold in the range between 0.5 and 0.3. \textbf{b, upper panel}: comparison for ’Pt19O2’ fragment in the 5x5x3 and 5x5x5 atomistic models; \textbf{b, lower panel}: in the narrower range between 0.2 and 0.1 AVAS threshold the behaviour of MP2 energies shows that: 1) the size effect in the energies, with the larger as a better model; 2) an AVAS threshold=0.18 is the better possible choice, since the corresponding MP2 energy is closer to the DFT reference value (light$-$red dash$-$dotted line).}\label{fig:density_diff2}
\end{figure}

\pagebreak
 
\section{Pure platinum VQE and VQE+ NEVPT2 H1-E noisy emulator experiments}
 
\begin{figure}[h]%
\centering
\includegraphics[width=\textwidth]{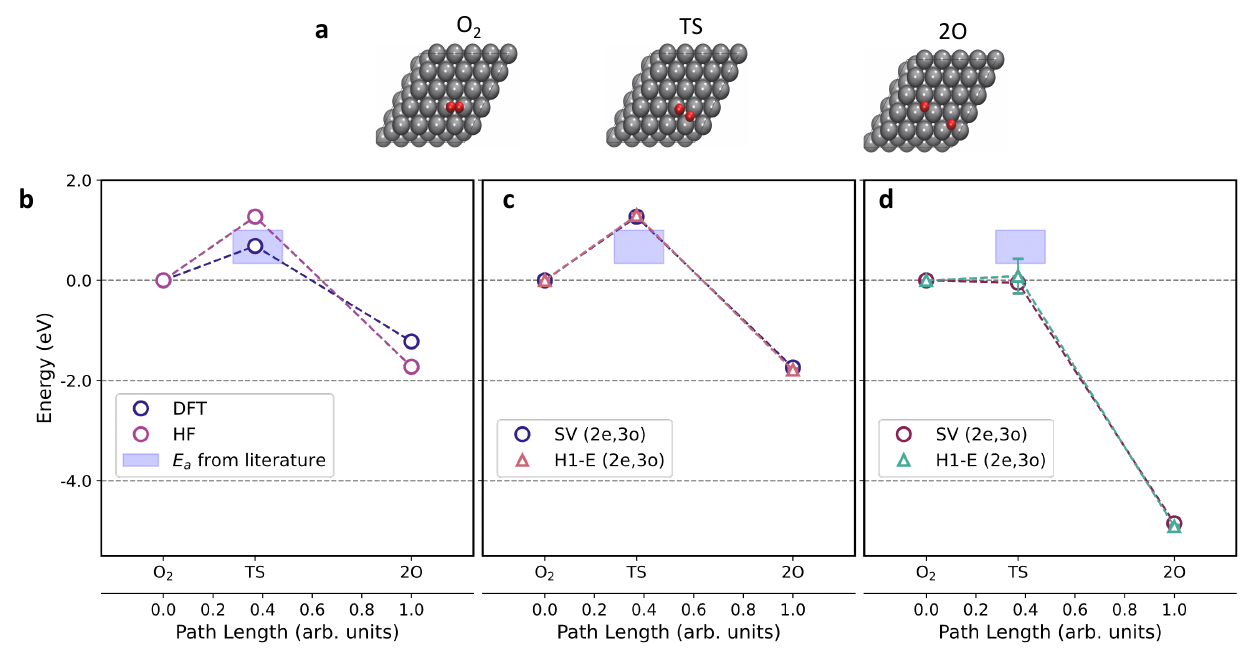}
\caption{\textbf{a}, Atomistic models for reactant (O$_2$, transition state (TS) and product (2O) for the dissociation of O$_2$ on platinum surface (see Fig. 3 main text). \textbf{b,c,d}, Energy profiles along the NEB reaction path (see Fig. 3b, in the main text):
\textbf{a}, the mean field calculations are DFT (in purple) and HF (in violet); \textbf{b}, energies computed using our quantum computing algorithm based on VQE for the SV (in violet) and the resulting measurement from the H1-E noisy emulator(in light-red); \textbf{c}, energies computed using our quantum computing algorithm based on VQE and corrected by including dynamic correlation (VQE+NEVPT2 energies) for the SV (in violet) and the resulting measurement from the H1-E noisy emulator (in green).  The error bars are calculated as a standard deviation around the mean obtained by bootstrapping 100k in 10 batches each for Quantinuum H1-E nosy emulator measurements. Activation energy $E_a$ available from literature \textcolor{blue}{Montemore et al. 2018} depicted as light-blue area.}\label{fig:density_diff3}
\end{figure}

 \pagebreak

 \section{Pt/Co slabs: Optimised atomistic geometries}

\begin{figure}[h]%
\centering
\includegraphics[width=0.9\textwidth]{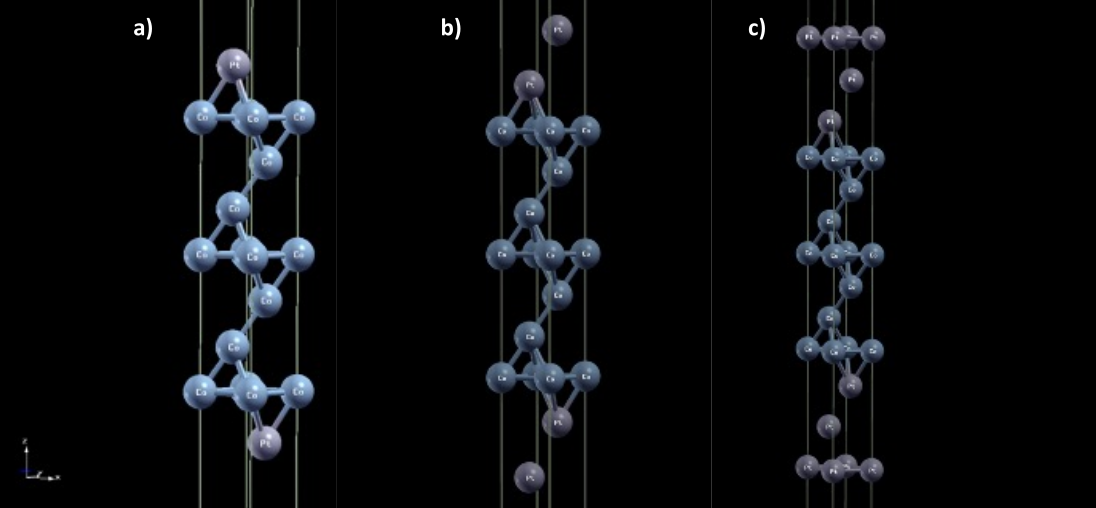}
\caption{Geometrically optimised Pt/Co slab atomistic structures with one (\textbf{a} panel), two (\textbf{b} panel) and three (\textbf{c} panel) Pt layers on top of a 7-layer cobalt core.}\label{fig:1_2_3ML_PtCo}
\end{figure}

 \pagebreak


\section{Details of Pt/Co slabs optimisation}

\begin{table}[h]
\begin{center}
\caption{Summary of the structural properties and energetics of 1, 2 a 3ML Pt capped Co slabs without surface relaxation. Notice that the total energy difference with plus sign signifies that the energy of two dissociated O atoms is higher than the energy of the O$_2$ molecule on the surface.}
\begin{tabular}{c|c|c|c}
\toprule
& 1 ML Pt & 2 ML Pt & 3 ML Pt   \\
\midrule
O-O dist.(\AA)               & 1.23    & 1.23    & 1.23      \\
O-Pt dist. (\AA)              & 3.32    & 3.41    & 3.37      \\
%m O    (mB)           & 0.561   & 0.559   & 0.553    \\
$\Delta E$ (Ry w.r.t.   O$_2$) & $+0.0841$  & $+0.11134$ & $+0.0501$  \\ 
\bottomrule
\end{tabular}
\end{center}
\end{table}

\begin{table}[h]
\begin{center}
\caption{Summary of the structural optimization for Pt/Co/Pt slabs with 1, 2 and 3 ML Pt capping. Reported are the inter-layer distances in \AA. Pure Pt bulk inter-layer distance amounts to $\approx 2.22$\AA.}
   \label{tabSslabs}
   \begin{tabular}{c|c|c|c}
\toprule 
distance (\AA)  & 1ML Pt & 2ML Pt & 3ML Pt\\

\midrule 
Pt-Pt &    -    &   -   & 2.6083\\
Pt-Pt &    -    & 2.6736& 2.5998\\
Pt-Co & 2.2418  & 2.2197& 2.2303\\
Co-Co & 1.9568  & 1.9727& 1.9589\\
Co-Co & 2.0125  & 2.0121& 2.0164\\
Co-Co & 1.9989  & 2.0023& 1.9997\\
Co-Co & 1.9989  & 2.0023& 1.9997\\
Co-Co & 2.0125  & 2.0121& 2.0164\\
Co-Co & 1.9568  & 1.9727& 1.9589\\
Pt-Co & 2.2418  & 2.2197& 2.2303\\
Pt-Pt &    -    & 2.6736& 2.5998\\
Pt-Pt &    -    &   -   & 2.6083\\
\bottomrule
   \end{tabular}
\end{center}
\end{table}

\begin{table}[h]
\begin{center}
\caption{Summary of the magnetic polarization for Pt/Co/Pt slabs with 1, 2 and 3 ML Pt capping. Reported are the Pt (Co) magnetic moments in Bohr magneton units per atom in each layer.}
\label{tabSmag}
   \begin{tabular}{c|c|c|c}
\toprule 
  & 1ML Pt ($\mu_B$) & 2ML Pt ($\mu_B$) & 3ML Pt ($\mu_B$) \\
\midrule 
Pt  &  -    &  -    &-0.0168\\
Pt  &  -    & 0.0472& 0.0037\\
Pt  & 0.1738& 0.1712& 0.1275\\
Co  & 1.8255& 1.8279& 1.7760\\
Co  & 1.6999& 1.7394& 1.7433\\
Co  & 1.7385& 1.7666& 1.7534\\
Co  & 1.7071& 1.7430& 1.7474\\
Co  & 1.7385& 1.7666& 1.7534\\
Co  & 1.6999& 1.7394& 1.7433\\
Co  & 1.8255& 1.8279& 1.7760\\
Pt  & 0.1738& 0.1712& 0.1275\\
Pt  &  -    & 0.0472& 0.0037\\
Pt  &  -    &  -    &-0.0168\\
\bottomrule
   \end{tabular}
\end{center}
\end{table}

\pagebreak

\section{AVAS VQE$_1$ for Pt/Co}

\begin{figure*}[h]
    \centering
    \includegraphics[width=\textwidth]{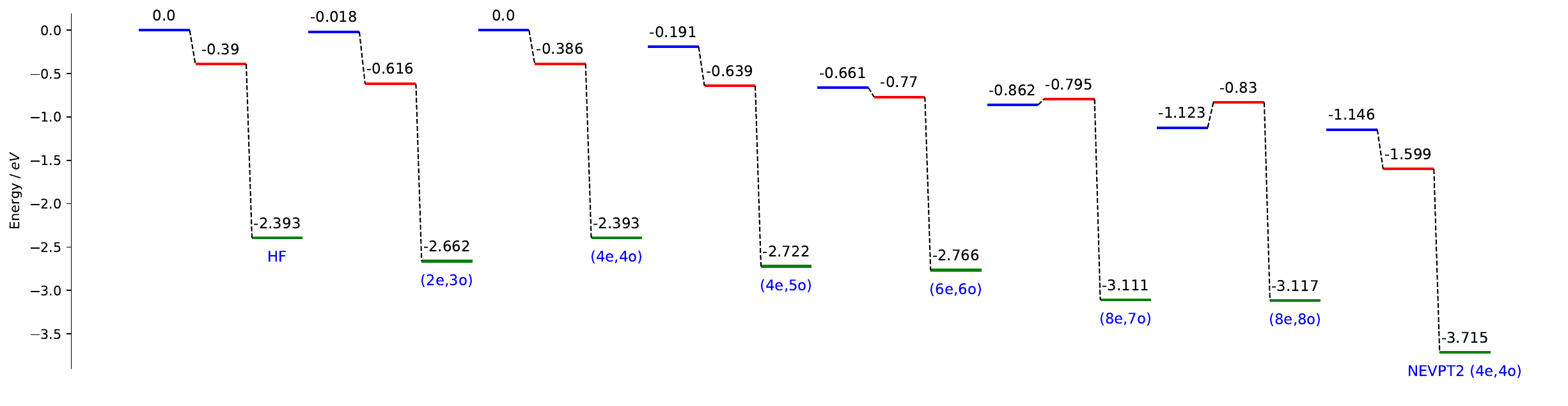}
    \caption{VQE state vector energy levels for all the states at different active space configurations and for Hartree-Fock, VQE and VQE+QRDM\_NEVPT2 methods. HF stands for Hartree-Fock. The energies are shifted with respect to the R state energy at the Hartree-Fock level. Blue, red and green represent O$_2$ (R), TS, 2O (P), respectively adsorbed on Pt/Co(111).}
    \label{fig:VQE}
\end{figure*}
%\newpage
%\bibliography{references}% common bib file